\documentclass[useAMS,onecolumn,usenatbib]{mn2e}

\usepackage{graphicx}
\usepackage{times}
\usepackage{amssymb}

\title[Fractal features in accretion discs]
{Fractal features in accretion discs} 
\author[Roy and Ray]
{Nirupam Roy$^{1}$\thanks{nirupam@ncra.tifr.res.in}
and Arnab K. Ray$^{2}$\thanks{akr@hbcse.tifr.res.in}\\
$^{1}$National Centre for Radio Astrophysics, Tata Institute of
Fundamental Research, Post Bag 3, Ganeshkhind, Pune 411007, India\\
$^{2}$Homi Bhabha Centre for Science Education, Tata Institute of 
Fundamental Research, V. N. Purav Marg, Mankhurd, Mumbai 400088, India}

\begin{document}


\maketitle

\label{firstpage}

\begin{abstract}
Fractal concepts have been introduced in the accretion disc as a 
new feature. Due to the fractal nature of the flow, its continuity 
condition undergoes modifications. The conserved stationary fractal
flow admits only saddle points and centre-type points in its phase 
portrait. Completely analytical solutions of the equilibrium point 
conditions indicate that the fractal properties enable the flow to 
behave like an effective continuum of lesser density, and facilitates
the generation of transonicity. However, strongly fractal flows inhibit 
multitransonicity from developing. The mass accretion rate exhibits
a fractal scaling behaviour, and the entire fractal accretion disc 
is stable under linearised dynamic perturbations. 
\end{abstract}

\begin{keywords}
accretion, accretion discs -- black hole physics -- hydrodynamics
\end{keywords}

\section{Introduction}
\label{sec1}

Accretion processes are understood to be the non-self-gravitating flow 
of a compressible astrophysical fluid, driven by the external gravitational 
field of a massive astrophysical object, like an ordinary star or a 
compact object or, what is perhaps of greatest interest in the context
of accretion studies, a black hole. The matter falling 
into the potential well of an accretor is commonly modelled in terms 
of a fluid continuum, and standard fluid dynamical equations, 
customized in the Newtonian framework of space and time, suffice 
well to provide a satisfactory description of the entire accretion 
process~\citep{fkr02}. While this be so, in recent times fractal modelling
of astrophysical accretion is generating some interest~\citep{roy07,rnr07}. 
One astrophysical material that ought readily to lend itself to a fractal
approach --- as opposed to a continuum approach --- is the interstellar
medium (ISM). It is recognised that the ISM is not entirely to be
seen as a fluid continuum, and for many purposes the ISM is believed 
to possess a self-similar hierarchical structure over several orders 
of magnitude in scale~\citep{lars,falg92,heith}.
Direct H~{\sc i} absorption observations and interstellar scintillation
measurements suggest that the structure extends down to a scale of
$10\, \mathrm{au}$~\citep{crov,lang,fais} and possibly even to
sub-$\mathrm{au}$ scales~\citep{hill}.
Numerous theories have attempted to explain the origin, evolution
and mass distribution of these clouds and it has been established,
from both observations~\citep{elme} and numerical
simulations~\citep{burk,kless,seme}, that the interstellar medium
has a clumpy hierarchical self-similar structure with a fractal
dimension in three-dimensional space. The main reason for this is 
still not properly understood, but it can be the consequence of an 
underlying fractal geometry that may arise due to turbulent processes 
in the medium.

A comprehensive mathematical description of an accreting system 
--- either a fluid continuum or a fractal structure --- will 
necessarily have to fall within the broad domain of nonlinear dynamics.
This is the principal basis of the analytical methods adopted for 
this study. In the context of spherically symmetric accretion, 
a previous work furnished a self-consistent description of the 
hydrodynamics in a fractal medium~\citep{roy07}. A similar description 
has been provided in this work for the case of a conserved rotational 
flow, which, from a general fluid dynamical perspective, has become 
quite thoroughly understood~\citep{sc81}. In the context of accretion 
studies particularly, on many occasions such flows are devised to be 
an inviscid sub-Keplerian compressible flow, and as such, this model 
has become well-established in accretion-related literature by 
now~\citep{az81,fuk87,
c89,nf89,skc90,ky94,yk95,par96,msc96,skc96,lyyy97,das02,dpm03,ray03,
bdw04,das04,abd06,dbd06,crd06,gkrd07}. As a first step, 
the physical processes in a fractal medium have been analysed by 
fractional integration and differentiation~\citep[][and references 
therein]{zas}. To do so, the fractal medium has had to be replaced
by a continuous medium and the integrals on the network of the fractal
medium has had to be approximated by fractional integrals~\citep{ren}.
The interpretation of fractional integration is connected with
fractional mass dimension~\citep{mand}. Fractional integrals can
be considered as integrals over fractional dimension space within
a numerical factor~\citep{tara}. This numerical factor has been
chosen suitably to get the right dimension of any physical parameter
and to derive the standard expression in the continuum limit. The
isotropic and homogeneous nature of dimensionality, in the context
of an axisymmetric flow, has also been incorporated properly. 

Having established the hydrodynamic equivalence, a standard 
mathematical path 
has been followed to set up the steady fractal flow like a simple 
first-order autonomous dynamical system~\citep{stro,js99} and 
investigate the resulting critical properties, especially the 
transonic character of the flow. A general prediction that can 
be made about the conserved steady fractal disc flow is that the 
equilibrium conditions will only permit the existence of saddle
points and centre-type points in the phase portrait of the flow. 
Fully analytical 
methods have been employed to identify the exact critical point
coordinates, under a pseudo-Newtonian prescription to study the 
infall process. Two very significant effects of fractal properties
can be discerned in consequence of this. The first is that fractal 
accretion flows assume the properties of an equivalent continuum, 
albeit one with a more diluted effective local flow density, a result 
that is also known to be true for the case of spherically symmetric 
flows~\citep{rnr07}{\footnote{In the context of fractals, it would 
be interesting to note a similar equivalence between the discrete 
model of diffusion-limited aggregation and the continuum dendritic 
growth model~\citep{witsan}.}}.
The second effect is that fractal properties have an inhibitory 
influence on the multitransonic character that is commonly 
associated with these flows~\citep{c89,skc90,skc96,das02,das04}. 
Taken together, the overall effect
of strongly fractal features in a disc flow is to lend it the 
appearance of the spherically symmetric monotransonic 
flow --- the paradigmatic~\citet{bon52} solution. 

In the Newtonian limit the mass accretion rate in a fractal flow 
exhibits a scaling behaviour that is governed by the fractal 
dimension of the flow. This scaling behaviour may have many 
interesting astrophysical implications. One particular case, that 
of accretion processes onto a pre-main-sequence (PMS) star, should 
have a close connection with this fractal scaling relation. 
It has also been shown that the 
stability of fractal flows under linearised time-dependent 
perturbations, is governed by the same equation pertaining 
to a standard perturbation scheme that has been used earlier 
for studying continuum sub-Keplerian disc 
flows~\citep{ray03,crd06,rbcqg}. 
Given that the boundary conditions do not vary between the continuum 
case and the fractal case, all steady global solutions (transonic 
or otherwise) are, therefore, stable. 
Since fractal disc flows tend to acquire the character of spherically 
symmetric flows, it has become possible to invoke the analogy
of the non-perturbative and dynamic evolution of the spherically
symmetric flow towards its transonic 
state~\citep{rb02,rnr07}, and suggest 
that in a fractal disc, the drive towards transonicity would also 
occur under identical guiding principles. 

It would be important to note that the entire analytical treatment 
has been carried out in the pseudo-Newtonian formalism, with the 
mathematical results expressed in terms of a general potential,
whose resulting force field would drive the accretion process. 
So most of the conclusions reached here will remain independent
of the choice of a potential. However, the specific numerical 
plots which have been presented here, have all been derived by 
using the~\citet{pw80} potential for a polytropic gas flow. In 
spite of this, the general claims made on the basis of these 
particular numerical results will suffer no qualitative alteration, 
under the choice of any other pseudo-Newtonian 
potential~\citep{nw91,abn96}, or for an isothermal gas flow~\citep{dpm03}. 

\section{The hydrodynamic equations for axisymmetric fractal accretion}
\label{sec2}

Considering a thin accretion disc that has density fluctuations over 
a range of length scales, it should be possible to propose that these 
density fluctuations or ``clumpiness" can be approximated as a fractal 
structure along the radial direction in the disc. This will be especially
true in the thin-disc regime, where there is not much scope for 
significant variations to occur in the vertical direction~\citep{fkr02}. 
The integrals on this network of fractals 
can consequently be approximated by fractional integrals~\citep{ren}, 
and the interpretation of this fractional integration can be connected 
to the fractional mass dimension. This approximation has already been 
used to describe the fractal structure of spherically symmetrical accretion 
by~\citet{roy07}. Likewise, a similar transformation of the infinitesimal 
length of the ``clumpy" accretion disc is being employed here to describe 
it in terms of an equivalent ``fractional continuous" disc. The fractional 
infinitesimal length for such a structure will be given by
\begin{equation}
\label{length}
{\mathrm{d}}\overline{r} = 
\left(\frac{r}{l_{\mathrm c}}\right)^{\Delta -1}{\mathrm d}r \,, 
\end{equation}
with $l_{\mathrm c}$ being the inner length scale of the structure, 
representing the physical scale of length at which the density is 
correlated. The fractal dimension of the accretion disc is given as 
$2\Delta$. Hence, the mass enclosed in a thin disc of density, $\rho$, 
radius, $r$, and thickness, $H$, will be
\begin{equation}
\label{md}
M_D = \int_0^r \int_{-H/2}^{H/2} 2\pi\rho\overline{r}\,
{\mathrm d}\overline{r}\,{\mathrm d}z \sim \rho H r^{2\Delta} \,, 
\end{equation}
which, quite obviously, will give the standard expression in the 
limit $\Delta \longrightarrow 1$.

Since there is no flow in the $z$ direction (perpendicular to the plane 
of the disc), the condition of hydrostatic equilibrium~\citep{fkr02}, 
by balancing the 
pressure gradient and the gravitational force acting on a fluid element
in the $r$--$\phi$--$z$ space, will give
\begin{equation}
\label{hybal}
-\frac{\partial}{\partial z}\left( P\overline{r}\,
{\mathrm d}{\overline{r}}\,
{\mathrm{d}}\phi \right){\mathrm d}z \simeq 
\Phi^{\prime}\rho \left(\frac{z}{r}\right)\overline{r}\,
{\mathrm d}{\overline{r}}\,
{\mathrm{d}}\phi \,{\mathrm d}z \,,
\end{equation}
in which $P$ is the gas pressure, and 
$\Phi$ is the gravitational potential of the central accretor that 
drives the flow (with the prime denoting the spatial derivative of the 
potential). In the case of stellar accretion, the flow is driven by the 
Newtonian potential, $\Phi (r)=-GM/r$. 
On the other hand, frequently in studies 
of black hole accretion, it becomes convenient to use an {\em effective} 
pseudo-Newtonian potential that imitates general relativistic effects in 
the Newtonian construct of space and time~\citep{pw80,nw91,abn96}. 

Along with the definition of a polytropic equation of state~\citep{sc39}, 
$P=K\rho^\gamma$ (with the polytropic exponent, $\gamma$, having the 
range $1 \le \gamma \le 5/3$), equation~(\ref{hybal}) can be simplified 
to give the condition for the thin-disc approximation~\citep{fkr02} as
\begin{equation}
\label{tdcon}
H \simeq c_{\mathrm s} 
\left(\frac{r}{\gamma\Phi^{\prime}}\right)^{1/2} \,,
\end{equation}
in which the local speed of sound, $c_{\mathrm s}$, has been defined 
by $c_{\mathrm s}^2 = \partial P/\partial\rho$. This definition, as 
well as the form of $H$ in equation~(\ref{tdcon}), will be instrumental 
in determining the mass density balance (the continuity condition) 
in a conserved fractal accretion disc. For an infinitesimal volume of the 
fractal medium, ${\mathrm d}V$, this balance will be given by
\begin{equation}
\label{balan}
\frac{\partial}{\partial t}\left(\rho \,{\mathrm d}\overline{V}\right) 
= -\frac{\partial}{\partial r}\left(\rho v\frac{{\mathrm d}
\overline{V}}{{\mathrm d}r}\right){\mathrm d}r \,.
\end{equation}
Making note of the connection that 
${\mathrm d}\overline{V}=H\overline{r}\,{\mathrm d}\overline{r}\,
{\mathrm d}\phi$ for the axisymmetric thin disc flow, 
a more informative form of equation~(\ref{balan})
can be derived as 
\begin{equation}
\label{sigcon}
\frac{\partial\Sigma}{\partial t} + \frac{1}{r^{2\Delta -1}}
\frac{\partial}{\partial r}\left(\Sigma vr^{2\Delta -1}\right)=0 \,,
\end{equation}
in which $\Sigma$ is the surface density, defined by
$\Sigma \simeq \rho H$~\citep{fkr02}. 
Using the thin-disc approximation, as given by equation~(\ref{tdcon}), 
one can recast equation~(\ref{sigcon}) further as 
\begin{equation}
\label{rhocon}
\frac{\partial}{\partial t}\left[\rho^{(\gamma+1)/2}\right] + 
\frac{\sqrt{\Phi^\prime}}{r^{\sigma}}
\frac{\partial}{\partial r}
\left[\rho^{(\gamma+1)/2}v \frac{r^{\sigma}}
{\sqrt{\Phi^\prime}}\right] = 0 \,,
\end{equation}
where $\sigma =2\Delta -(1/2)$. The foregoing expression gives the  
mass density balance equation (the continuity equation) for the 
axisymmetric fractal flow. 

Similarly, in the infinitesimal volume, ${\mathrm d}V$, the balance 
of momentum density will imply 
\begin{equation}
\label{mombal}
\frac{\mathrm d}{{\mathrm d}t}\left(\rho {\mathbf v} \,
{\mathrm d}\overline{V}\right)={\mathbf F}_{\mathrm g} 
+{\mathbf F}_{\mathrm{cf}} +{\mathbf F}_{\mathrm p} \,,
\end{equation}
with ${\mathbf v}$ being the velocity vector, ${\mathbf F}_{\mathrm g}$ 
and ${\mathbf F}_{\mathrm{cf}}$ being, respectively, the gravitational 
and centrifugal forces acting on the mass contained in the infinitesimal 
volume, ${\mathrm d}V$, and ${\mathbf F}_{\mathrm p}$ being the total 
surface force due to the pressure acting on the full surface bounding the 
volume, ${\mathrm d}V$. The first two forces have radial components only,
and their magnitudes are given by
\begin{equation}
\label{radgrav}
F_{\mathrm g} = -\Phi^{\prime} \rho \,{\mathrm d}\overline{V} 
\end{equation}
and
\begin{equation}
\label{radcent}
F_{\mathrm{cf}} = \frac{\lambda^2}{r^3} \rho \,{\mathrm d}\overline{V} \,,
\end{equation}
respectively, with $\lambda$ in the latter being the constant specific 
angular momentum of the conserved disc flow~\citep{c89,skc90,skc96}. On the 
other hand, ${\mathbf F}_{\mathrm p}$ has $r$, $\phi$ and $z$ components. 
However, only the radial component is relevant for the axisymmetric 
thin-disc flow. Any volume element, ${\mathrm d}V$, experiences a force,  
$-{\mathrm d}V({\mathbf \nabla}P)$, due to the pressure exerted on it
by the surrounding medium~\citep{ll87}. Translating this effect onto an 
infinitesimal volume element of the fractal medium, the magnitude of the 
radial component of the force, ${\mathbf F}_{\mathrm p}$, can be set 
down as 
\begin{equation}
\label{radpres}
F_{\mathrm p} = -\frac{1}{\rho}\frac{\partial P}{\partial r} 
 \rho \,{\mathrm d}\overline{V} \,.
\end{equation}

Now going back to the left-hand side of equation~(\ref{mombal}), the 
total change of radial momentum is extracted as 
\begin{equation}
\label{lhsmombal}
\frac{\mathrm d}{{\mathrm d}t}\left(\rho {\mathbf v} \,
{\mathrm d}\overline{V}\right) = \left(\frac{\partial v}{\partial t} 
+v\frac{\partial v}{\partial r}\right)\rho \,{\mathrm d}\overline{V} \,,
\end{equation}
a result that could be derived by invoking the continuity condition, 
as it is given by equation~(\ref{balan}). And so, making use of 
equations~(\ref{radgrav}), (\ref{radcent}), (\ref{radpres}) 
and~(\ref{lhsmombal}) in equation~(\ref{mombal}), it becomes possible
to obtain the final momentum balance condition for the conserved 
fractal disc flow as  
\begin{equation}
\label{fulleuler}
\frac{\partial v}{\partial t} + v\frac{\partial v}{\partial r} 
+ \frac{1}{\rho}\frac{\partial P}{\partial r} + \Phi^{\prime}(r) 
- \frac{\lambda^2}{r^3} = 0 \,,
\end{equation}
which actually gives a local conservation law and, as expected, has 
exactly the same form as that of the continuous medium. The fractal
flow can now be described completely by equations~(\ref{rhocon}) 
and~(\ref{fulleuler}), both of which describe the dynamics of the
fields $v(r,t)$ and $\rho (r,t)$, with the pressure, $P$, having 
been defined as a function of $\rho$. 

\section{The stationary fractal flow and its fixed points}
\label{sec3}

It is a standard practice to consider a steady flow while studying
accreting systems~\citep{fkr02}. 
The two corresponding stationary equations which 
determine the drift in the radial direction, can be obtained from 
equations~(\ref{rhocon}) and~(\ref{fulleuler}) as 
\begin{equation}
\label{con}
\frac{\mathrm{d}}{\mathrm{d}r}
\left[\rho^{(\gamma+1)/2}v \frac{r^{\sigma}}
{\sqrt{\Phi^\prime}}\right] = 0 
\end{equation}
and 
\begin{equation}
\label{euler}
v \frac{\mathrm{d}v}{\mathrm{d}r}
+ \frac{1}{\rho}\frac{\mathrm{d}P}{\mathrm{d}r}
+ \Phi^{\prime}(r) - \frac{\lambda^2}{r^3} = 0 \,,
\end{equation}
respectively. The pressure, $P$, is prescribed by an equation of 
state for the flow~\citep{sc39}. As a general polytropic it is, 
as usual, given as $P=K \rho^{\gamma}$,
while for an isothermal flow the pressure is given by
$P= \rho {\kappa}T/\mu m_{\mathrm{H}}$, in all of which,
$K$ is a measure of the entropy in the flow, $\gamma$ is the
polytropic exponent, $\kappa$ is Boltzmann's constant, $T$ is the
constant temperature, $m_{\mathrm{H}}$ is the mass of a hydrogen
atom and $\mu$ is the reduced mass, respectively. 
Since the local speed of sound is defined as
$c_{\mathrm{s}} = (\partial P/\partial \rho)^{1/2}$, the way
in which $P$ has been prescribed~\citep{fkr02} will affect the 
transonic features of the flow, because transonicity, and all 
its related critical aspects, are measured by scaling the bulk 
flow velocity with respect to the local speed of sound. So the 
exact form of $P$ has distinctive consequences, and in what follows,
the flow properties will be taken up separately for the two cases,
i.e. polytropic and isothermal.

\subsection{Polytropic flows}
\label{subsec31}

With the polytropic relation specified for $P$, it is a
straightforward exercise to set down in terms of the speed of sound, 
$c_{\mathrm{s}}$, a first integral of equation~(\ref{euler}) as,
\begin{equation}
\label{eupol1st}
\frac{v^2}{2} + n c_{\mathrm{s}}^2 + \Phi (r)
+ \frac{\lambda^2}{2 r^2} = \mathcal{E} \,,
\end{equation}
in which $n=(\gamma -1)^{-1}$ and the integration constant
$\mathcal{E}$ is the Bernoulli constant. The first integral of 
equation~(\ref{con}) could similarly be obtained as 
\begin{equation}
\label{conpol1st}
c_{\mathrm{s}}^{2(2n +1)} \frac{v^2 r^{2 \sigma}}{\Phi^{\prime}}
= \frac{\gamma}{4 \pi^2} \dot{\mathcal{M}}^2 \,,
\end{equation}
where $\dot{\mathcal{M}} = (\gamma K)^n \dot{m}$~\citep{skc90,skc96}
with $\dot{m}$, an integration constant itself, being physically the
matter flow rate.

To obtain the critical points of the flow, it should be necessary
first to differentiate both equations~(\ref{eupol1st}) 
and~(\ref{conpol1st}), and then, on combining the two resulting 
expressions, to arrive at
\begin{equation}
\label{dvdrpol}
\left(v^2 - \beta^2 c_{\mathrm{s}}^2 \right)
\frac{\mathrm{d}}{\mathrm{d}r}(v^2) = \frac{2 v^2}{r}
\left[ \frac{\lambda^2}{r^2} - r \Phi^{\prime}
+ \frac{1}{2}\beta^2 c_{\mathrm{s}}^2
\left(2\sigma - r \frac{\Phi^{\prime \prime}}
{\Phi^{\prime}}\right)\right] \,,
\end{equation}
with $\beta^2 = 2(\gamma +1)^{-1}$. The critical points of the flow
will be given by the condition that the entire right-hand side of
equation~(\ref{dvdrpol}) will vanish along with the coefficient of
${\mathrm{d}}(v^2)/{\mathrm{d}r}$. Explicitly written down,
following some rearrangement of terms, this will give the two
critical point conditions as
\begin{equation}
\label{critconpol}
v_{\mathrm{c}}^2 = \beta^2 c_{\mathrm{sc}}^2
= 2\left[r_{\mathrm{c}} \Phi^{\prime}(r_{\mathrm{c}})
- \frac{\lambda^2}{r_{\mathrm{c}}^2}\right]\left[2\sigma -r_{\mathrm{c}}
\frac{\Phi^{\prime \prime}(r_{\mathrm{c}})}{\Phi^{\prime}(r_{\mathrm{c}})}
\right]^{-1} \,,
\end{equation}
with the subscript ``${\mathrm{c}}$" labelling critical point values.

To fix the critical point coordinates, $v_{\mathrm{c}}$ and 
$r_{\mathrm{c}}$, in terms of the system constants, one would have 
to make use of the conditions given by equations~(\ref{critconpol}), 
along with equation~(\ref{eupol1st}), to obtain
\begin{equation}
\label{efixcrit}
\frac{2 \gamma}{\gamma -1}
\left[r_{\mathrm{c}} \Phi^{\prime}(r_{\mathrm{c}})
-\frac{\lambda^2}{r_{\mathrm{c}}^2}\right]\left[2\sigma -r_{\mathrm{c}}
\frac{\Phi^{\prime \prime}(r_{\mathrm{c}})}{\Phi^{\prime}(r_{\mathrm{c}})}
\right]^{-1} + \Phi (r_{\mathrm{c}})
+ \frac{\lambda^2}{2 r_{\mathrm{c}}^2} = \mathcal{E} \,,
\end{equation}
from which it is easy to see that solutions of $r_{\mathrm{c}}$ may be 
obtained in terms of $\gamma$, $\lambda$, $\sigma$ (or $\Delta$) and 
$\mathcal{E}$ only, i.e.
$r_{\mathrm{c}}=f_1(\gamma, \lambda, \sigma, \mathcal{E})$.
Alternatively, $r_{\mathrm{c}}$
could be fixed in terms of $\gamma$, $\lambda$, $\sigma$ 
and $\dot{\mathcal{M}}$. By making use of the critical point conditions 
in equation~(\ref{conpol1st}), one could write
\begin{equation}
\label{dotmfix}
\frac{4 \pi^2 \beta^2 r_{\mathrm{c}}^{2\sigma}}
{\gamma\Phi^{\prime}(r_{\mathrm{c}})}
\left[ \frac{2}{\beta^2}
\left\{r_{\mathrm{c}} \Phi^{\prime}(r_{\mathrm{c}})
- \frac{\lambda^2}{r_{\mathrm{c}}^2}\right\}\left\{2\sigma -r_{\mathrm{c}}
\frac{\Phi^{\prime \prime}(r_{\mathrm{c}})}{\Phi^{\prime}(r_{\mathrm{c}})}
\right\}^{-1}\right]^{2(n +1)} = {\dot{\mathcal{M}}}^2 \,,
\end{equation}
whose obvious implication is that the dependence of $r_{\mathrm{c}}$
will be given as 
$r_{\mathrm{c}}= f_2(\gamma, \lambda, \sigma, \dot{\mathcal{M}})$.

The slope of the continuous solutions which could possibly pass through
the critical points are to be obtained by applying the L'Hospital rule
on equation~(\ref{dvdrpol}), at the critical points. This will give a
quadratic equation for the slope of the stationary solutions at the 
critical points themselves in the $r$ --- $v^2$ phase portrait. The 
resulting expression will read as
\begin{equation}
\label{lhosp}
\left[\frac{\mathrm{d}}{\mathrm{d}r}(v^2)
{\bigg{\vert}}_{\mathrm c}\right]^2
+ {\mathcal{Z}}_1 
\left[\frac{\mathrm{d}}{\mathrm{d}r}(v^2)
{\bigg{\vert}}_{\mathrm c}\right]
+ {\mathcal{Z}}_0 = 0 ,
\end{equation}
in which the constant coefficients, $\mathcal{Z}_1$
and $\mathcal{Z}_0$, are given by
\begin{displaymath}
\label{zed1}
{\mathcal Z}_1 =\frac{2}{\gamma}\left(\frac{\gamma -1}
{\gamma +1}\right) \frac{c_{\mathrm{sc}}^2}{r_{\mathrm c}}
\left[2\sigma - r_{\mathrm{c}}\frac{\Phi^{\prime \prime}
(r_{\mathrm{c}})}{\Phi^{\prime}(r_{\mathrm{c}})}\right] 
\end{displaymath}
and 
\begin{displaymath}
\label{zed0}
{\mathcal Z}_0 = \frac{c_{\mathrm{sc}}^2}{\gamma}
\left[\frac{6 \lambda^2}{r_{\mathrm c}^4} + 
2 \Phi^{\prime \prime}(r_{\mathrm{c}}) + \frac{2}{\gamma +1}
c_{\mathrm{sc}}^2 \left\{ \frac{2\sigma}{r_{\mathrm c}^2} +
\frac{\mathrm{d}}{\mathrm{d}r}
\left(\frac{\Phi^{\prime \prime}}{\Phi^{\prime}}
\right) {\bigg{\vert}}_{\mathrm c} \right\} + 
\left(\frac{\gamma -1}{\gamma +1}\right)
 \frac{v_{\mathrm{c}}^2}{r_{\mathrm c}^2}
\left \{2 \sigma - r_{\mathrm{c}}\frac{\Phi^{\prime \prime}
(r_{\mathrm{c}})}{\Phi^{\prime}(r_{\mathrm{c}})}\right \}^2
\right] \,,
\end{displaymath}
respectively. 

\subsection{Isothermal flows}
\label{subsec32}

For isothermal flows, a linear dependence between $P$ and $\rho$ is 
applied in equation~(\ref{euler}), as the appropriate equation of state. 
On doing so, the first integral of equation~(\ref{euler}) is given as
\begin{equation}
\label{euiso1st}
\frac{v^2}{2} + c_{\mathrm s}^2 \ln \rho + \Phi (r)
+ \frac{\lambda^2}{2 r^2} = \mathcal{C} \,,
\end{equation}
with $\mathcal{C}$ being a constant of integration. For flow solutions
which specifically decay out to zero at very large distances, the constant
$\mathcal{C}$ can be determined in terms of the ``ambient conditions" 
as $\mathcal{C} = c_{\mathrm s}^2 \ln \rho_\infty$. The thickness of 
the disc, $H$, is determined simply by setting $\gamma =1$ in 
equation~(\ref{tdcon}), and in a likewise manner the first integral 
of equation~(\ref{con}) will imply the continuity condition as
\begin{equation}
\label{coniso1st}
\frac{\rho^2 v^2 r^{2\sigma}}{\Phi^{\prime}}
= \frac{\dot{m}^2}{4 \pi^2 c_{\mathrm{s}}^2} \,.
\end{equation}

As has been done for polytropic flows,
both equations~(\ref{euiso1st}) and~(\ref{coniso1st}) are
to be differentiated and the results combined to give
\begin{equation}
\label{dvdriso}
\left(v^2 -  c_{\mathrm{s}}^2 \right)
\frac{\mathrm{d}}{\mathrm{d}r}(v^2) = \frac{2 v^2}{r}
\left[ \frac{\lambda^2}{r^2} - r \Phi^{\prime}
+ \frac{1}{2} c_{\mathrm{s}}^2
\left(2\sigma -r \frac{\Phi^{\prime \prime}}
{\Phi^{\prime}}\right)\right] \,,
\end{equation}
from which the critical point conditions are easily identified as
\begin{equation}
\label{critconiso}
v_{\mathrm{c}}^2 = c_{\mathrm{s}}^2
= 2\left[r_{\mathrm{c}} \Phi^{\prime}(r_{\mathrm{c}})
-\frac{\lambda^2}{r_{\mathrm{c}}^2}\right]\left[2\sigma -r_{\mathrm{c}}
\frac{\Phi^{\prime \prime}(r_{\mathrm{c}})}{\Phi^{\prime}(r_{\mathrm{c}})}
\right]^{-1} \,.
\end{equation}
In this isothermal system, the speed of sound, $c_{\mathrm{s}}$, is
a global constant, and so having arrived at the critical point
conditions, it should be easy to see that $r_{\mathrm{c}}$ and
$v_{\mathrm{c}}$ have already been fixed in terms of a global
constant of the system. The speed of sound can further be written
in terms of the temperature of the system as
$c_{\mathrm{s}} = \Theta T^{1/2}$, where
$\Theta = (\kappa/\mu m_{\mathrm H})^{1/2}$, and, therefore,
it should be entirely possible to give a functional dependence for
$r_{\mathrm{c}}$, as $r_{\mathrm{c}} = f_3(\lambda, \sigma, T)$.
The slope of the stationary solutions passing through the critical 
points in the $r$ --- $v^2$ phase portrait is obtained simply by 
setting $\gamma =1$ in equation~(\ref{lhosp}).

\section{The fractal accretion disc as a dynamical system} 
\label{sec4}

The equations governing the flow in an accreting system are all
fluid dynamical equations, and, in the kind of inviscid regime 
chosen for this study, fall under the general category of first-order 
nonlinear differential equations~\citep{js99}. However, there is 
no standard prescription for deriving rigorously analytical solutions 
of these equations. 
In this situation, while a numerical integration of the flow 
equations is most often the only recourse, an alternative approach 
could also be adopted by setting up the governing equations 
to form a standard first-order dynamical system~\citep{stro,js99}.
This is a very usual practice in general fluid dynamical
studies~\citep{bdp93}, and avoiding involved numerical processes, 
this approach allows one to derive significant physical insight 
about the behaviour of the flows. To take a first step towards 
this end, for the stationary polytropic flow, as it has been given 
by equation~(\ref{dvdrpol}),
it should be necessary to parametrise this equation and set up
a coupled autonomous first-order dynamical system as~\citep{stro,js99}
\begin{eqnarray}
\label{dynsys}
\frac{\mathrm{d}}{\mathrm{d}\tau}(v^2)&=& 2v^2 \left[
\frac{\lambda^2}{r^2} -r\Phi^{\prime} +\frac{1}{2}
\beta^2 c_{\mathrm{s}}^2\left(2\sigma -r\frac{\Phi^{\prime \prime}}
{\Phi^{\prime}} \right) \right] \nonumber \\
\frac{\mathrm{d}r}{\mathrm{d} \tau}&=& r \left(v^2 -
\beta^2 c_{\mathrm{s}}^2 \right) \,,
\end{eqnarray}
with $\tau$ being an arbitrary mathematical parameter. Apropos of
accretion studies, this kind of parametrisation
has been reported before~\citep{rb02,ap03,crd06,rbcqg,mrd07,gkrd07}.
Some earlier works in accretion had also made use of the general 
mathematical aspects of this approach~\citep{mkfo84,mc86,ak89}

The critical points have themselves been fixed in terms of the flow
constants. About these fixed point values, on applying a perturbation
prescription of the kind
$v^2 = v_{\mathrm{c}}^2 + \delta v^2$, $c_{\mathrm{s}}^2 =
c_{\mathrm{sc}}^2 + \delta c_{\mathrm{s}}^2$ and
$r = r_{\mathrm{c}} + \delta r$, one could derive a set of two
autonomous first-order linear differential equations in the
$\delta r$ --- $\delta v^2$ plane, with $\delta c_{\mathrm{s}}^2$ itself
having to be first expressed in terms of $\delta r$ and $\delta v^2$,
with the help of equation~(\ref{conpol1st}) as
\begin{equation}
\label{varsound}
\frac{\delta c_{\mathrm{s}}^2}{c_{\mathrm{sc}}^2} = - \frac{\gamma -1}
{\gamma + 1} \left[\frac{\delta v^2}{v_{\mathrm{c}}^2}
+ \left\{2\sigma -r_{\mathrm{c}}\frac{\Phi^{\prime \prime}(r_{\mathrm{c}})}
{\Phi^{\prime}(r_{\mathrm{c}})} \right \}
\frac{\delta r}{r_{\mathrm{c}}}\right] \,.
\end{equation}
The resulting coupled set of linear equations in $\delta r$ and
$\delta v^2$ will be given as
\begin{eqnarray}
\label{lindynsys}
\frac{1}{{2v_{\mathrm{c}}^2}}\frac{\mathrm{d}}{\mathrm{d}\tau}
(\delta v^2) &=&  \frac{\mathcal A}{2}\left(\frac{\gamma -1}{\gamma + 1}
\right) \delta v^2 - \left [\frac{2 \lambda^2}{r_{\mathrm{c}}^3} +
\Phi^{\prime}(r_{\mathrm{c}}) + r_{\mathrm{c}}\Phi^{\prime \prime}
(r_{\mathrm{c}}) + \frac{\beta^2}{2}
\frac{\Phi^{\prime \prime}(r_{\mathrm{c}})}{\Phi^{\prime}(r_{\mathrm{c}})}
{c_{\mathrm{sc}}^2} \mathcal{B} + \frac{\beta^2}{2}
\left(\frac{\gamma -1}{\gamma + 1} \right)
\frac{{c_{\mathrm{sc}}^2}}{r_{\mathrm{c}}}
{\mathcal A}^2 \right] \delta r \nonumber \\
\frac{1}{r_{\mathrm{c}}}\frac{\mathrm{d}}{\mathrm{d}\tau}(\delta r)
&=& \frac{2\gamma}{\gamma + 1} \delta v^2 - \mathcal{A} \left(
\frac{\gamma -1}{\gamma + 1} \right)
\frac{{v_{\mathrm{c}}^2}}{r_{\mathrm{c}}} \delta r \,,
\end{eqnarray}
in which
\begin{displaymath}
\label{coeffs}
\mathcal{A} = r_{\mathrm{c}}\frac{\Phi^{\prime \prime}(r_{\mathrm{c}})}
{\Phi^{\prime}(r_{\mathrm{c}})} - 2\sigma \,,  \qquad
\mathcal{B} = 1 + r_{\mathrm{c}}
\frac{\Phi^{\prime \prime \prime}(r_{\mathrm{c}})}
{\Phi^{\prime \prime}(r_{\mathrm{c}})}
- r_{\mathrm{c}}\frac{\Phi^{\prime \prime}(r_{\mathrm{c}})}
{\Phi^{\prime}(r_{\mathrm{c}})} \,.
\end{displaymath}
Trying solutions of the kind $\delta v^2 \sim \exp(\Omega \tau)$ and 
$\delta r \sim \exp(\Omega \tau)$ in equations~(\ref{lindynsys}), will
lead to an expression for the eigenvalues, $\Omega$, of the stability
matrix, as 
\begin{equation}
\label{eigen}
\Omega^2 = \frac{4 r_{\mathrm{c}}
\Phi^{\prime}(r_{\mathrm{c}})c_{\mathrm{sc}}^2}{(\gamma + 1)^2}
\left[ \left\{ \left(\gamma - 1 \right){\mathcal A} 
-2 \gamma \left(2\sigma +1 + {\mathcal A} \right) +2\gamma 
{\mathcal{B}}\left(1 + \frac{2\sigma}{\mathcal A}\right)\right\}
- \frac{\lambda^2}{\lambda_{\mathrm K}^2(r_{\mathrm{c}})}
\left \{4 \gamma + \left(\gamma - 1 \right){\mathcal A} + 2 \gamma
{\mathcal{B}}\left(1 +\frac{2\sigma}{\mathcal A}\right)\right\}\right] \,,
\end{equation}
where $\lambda_{\mathrm K}^2(r) = r^3 \Phi^{\prime}(r)$.

For isothermal flows, starting from equation~(\ref{dvdriso}), a 
similar expression for the related eigenvalues may also be derived, 
and it can be shown that this will correspond simply to the special
case of $\gamma =1$ in equation~(\ref{eigen}). 
Once a critical point coordinate, ($r_{\mathrm c}, v_{\mathrm c}^2$), 
has been ascertained, it becomes an easy task to find the nature of 
that critical point by using $r_{\mathrm{c}}$ in equation~(\ref{eigen}).
Since it has been discussed in Section~\ref{sec3} that $r_{\mathrm{c}}$
is a function of $\gamma$, $\lambda$, $\sigma$ and $T$ for isothermal 
flows, and a function of $\gamma$, $\lambda$, $\sigma$ and $\mathcal E$ 
(or $\dot{\mathcal M}$) for polytropic flows, it effectively implies 
that $\Omega^2$ can be expressed as a function of the flow parameters 
for either kind of flow. A generic conclusion that can be drawn about 
the critical points from the form of $\Omega^2$ in equation~(\ref{eigen}), 
is that for a conserved pseudo-Schwarzschild fractal disc flow, the only
admissible critical points will be saddle points and centre-type points.
For a saddle point, $\Omega^2 > 0$, while for a centre-type point,
$\Omega^2 < 0$. Once the behaviour of all the physically relevant
critical points has been understood in this way, a complete qualitative
picture of the flow solutions passing through these points (if they
are saddle points), or in the neighbourhood of these points (if they
are centre-type points), can be constructed, along with an impression
of the direction that these solutions can have in the phase portrait
of the flow~\citep{stro,js99}. An earlier study has shown how the critical 
point behaviour (and multitransonicity) varies with flow parameters 
like $\lambda$ and $\mathcal E$~\citep{crd06}. 
Similar aspects of the accretion disc will be considered now 
with respect to the fractal properties in the flow.  

\section{Multitransonicity in relation to fractal features} 
\label{sec5}

Multitransonicity in the flow is possible when there are more than 
one critical point through which a continuous solution will pass,
connecting the event horizon of the accretor to the outer boundary 
of the flow. Saddle points are most likely to fulfil this requirement,
and so multitransonicity would imply the existence of more than one
saddle point in the phase portrait of the flow. Disc flows, conserved 
or otherwise, have been known to be multitransonic~\citep{lt80,skc90,
skc96} with a variety of implications, most notably in regard to shock 
formation~\citep{skc90,skc96,das02}. 
So a clear knowledge of the number of critical points in the flow,
and their nature, is essential to understanding multitransonicity. 

For a general polytropic flow, 
the position of the critical points can be determined by extracting
the roots from either equation~(\ref{efixcrit}) or~(\ref{dotmfix}),
after prescribing an explicit functional form for the pseudo-Newtonian 
potential, $\Phi (r)$. Various prescriptions are used for $\Phi (r)$,
depending upon specific necessities~\citep{pw80,nw91,abn96}, but all 
of these potentials have the common purpose of describing general 
relativistic effects in a semi-Newtonian framework, by avoiding the 
difficulties of a purely general relativistic treatment. One such 
potential that is very regularly invoked in accretion-related 
literature, is due to~\citet{pw80}. 
It is given as
\begin{equation}
\label{pwpot}
\Phi (r) = - \frac{1}{2\left(r - 1\right)} \,,
\end{equation}
in which the radial distance, $r$, has been scaled by the Schwarzschild
radius, $r_{\mathrm g}$, which follows the standard definition of
$r_{\mathrm g} = 2GM/c^2$, with $M$ being the mass of the black hole, 
$G$ the universal gravitational constant, and $c$ the velocity of light 
in vacuum. All relevant velocities will be scaled with respect to $c$,
and for convenience one may set $G=c=M=1$. All derived quantities like 
energy and angular momentum will be scaled accordingly~\citep{das02}. 

Using the functional form of $\Phi$, given by equation~(\ref{pwpot}), 
in equation~(\ref{efixcrit}), will lead to a quartic polynomial in 
the standard form
\begin{equation}
\label{quartic}
r_{\mathrm c}^4 + 2A r_{\mathrm c}^3 + Br_{\mathrm c}^2 
+ 2Cr_{\mathrm c} + D =0 \,,
\end{equation}
in which
\begin{displaymath}
\label{coeffabpol}
A = \frac{1}{4 {\mathcal E}\left(\sigma +1\right)} \left[\sigma + 1
- \frac{\gamma}{\gamma -1} - 2 {\mathcal E} \left(2 \sigma +1\right)
\right] \,,  \qquad \qquad 
B = \frac{1}{2 {\mathcal E}\left(\sigma +1\right)} \left[\lambda^2
\left(\frac{2\gamma}{\gamma -1} - \sigma -1 \right) + \sigma \left(
2 {\mathcal E} -1 \right) \right] \,,
\end{displaymath}
\begin{displaymath}
\label{coeffcdpol}
C = \frac{\lambda^2}{4 {\mathcal E}\left(\sigma +1\right)}
\left[2\sigma +1 - \frac{4\gamma}{\gamma -1} \right] \,,
\qquad \qquad \qquad 
D = \frac{\lambda^2}{2 {\mathcal E}\left(\sigma +1\right)}
\left[ \frac{2\gamma}{\gamma -1} - \sigma \right] \,.
\end{displaymath}
It is not difficult to see that equation~(\ref{quartic}) will yield
four roots. These roots can be found completely analytically by using
Ferrari's method for solving quartic equations. In order to do so,
a term going like $(a r_{\mathrm c} + b)^2$ is to be added to both 
sides of equation~(\ref{quartic}), and then the resulting left hand 
side is required to be a perfect square in the form 
$(r_{\mathrm c}^2 + Ar_{\mathrm c} + \zeta)^2$, so that the full 
equation will be rendered as
$(r_{\mathrm c}^2+Ar_{\mathrm c}+\zeta)^2=(ar_{\mathrm c}+b)^2$.
This will give three conditions going as 
\begin{equation}
\label{condi}
a^2 = \frac{\left(A \zeta - C\right)^2}{\zeta^2 - D} \,, \qquad \qquad
b = \frac{A \zeta - C}{a} \,, \qquad \qquad \zeta^2 = D + b^2 \,.
\end{equation}
Eliminating $a$ and $b$, will deliver an auxiliary cubic equation in
$\zeta$, going as
\begin{equation}
\label{auxil}
2 \zeta^3 - B \zeta^2 + 2\left(AC - D\right)\zeta 
+ \left(BD -A^2 D -C^2 \right) = 0 \,,
\end{equation}
which, under the transformation, $\zeta = \eta +(B/6)$, can be reduced 
to the canonical form of the cubic equation, 
\begin{equation}
\label{etacube}
\eta^3 + P\eta + Q = 0 \,,
\end{equation}
with 
\begin{displaymath}
\label{peeque}
P = -\frac{B^2}{12} + \left(AC - D\right) \,, \qquad \qquad
Q = - \frac{B^3}{108} + \frac{B}{6}\left(AC -D\right)+ \frac{1}{2}
\left(BD -A^2 D -C^2\right) \,.
\end{displaymath}
Completely analytical solutions for the roots of 
equation~(\ref{etacube}) can be obtained by the application of 
the Cardano-Tartaglia-del Ferro method for solving cubic equations.
This will lead to the solution
\begin{equation}
\label{cardan}
\eta = \left(-\frac{Q}{2} + \sqrt{\mathcal D} \right)^{1/3}
+ \left(-\frac{Q}{2} - \sqrt{\mathcal D} \right)^{1/3} \,,
\end{equation}
with the discriminant, $\mathcal D$, having been defined by
\begin{equation}
\label{discri}
{\mathcal D} = \frac{Q^2}{4} + \frac{P^3}{27} \,.
\end{equation}
The sign of $\mathcal D$ is crucial here. If ${\mathcal D} >0$, 
then there will be only one real root of $\eta$ given directly by 
equation~(\ref{cardan}). On the other hand, if ${\mathcal D} <0$, 
then there will be three real roots of $\eta$, all of which,
under a new definition,
\begin{equation}
\label{theta}
\vartheta = \arccos \left[\frac{-Q/2}
{\sqrt{-\left(P/3\right)^3}}\right] \,,
\end{equation}
can be expressed in a slightly modified form as 
\begin{equation}
\label{etatheta}
\eta_j = 2 \sqrt{\frac{-P}{3}} \cos \left[\frac{\vartheta + 2 \pi
\left(j -1 \right)}{3} \right] \,,
\end{equation}
with the label $j$ taking the values $j=1,2,3$, respectively,
for the three distinct roots.

\begin{figure}
\begin{center}
\includegraphics[scale=0.75, angle=0.0]{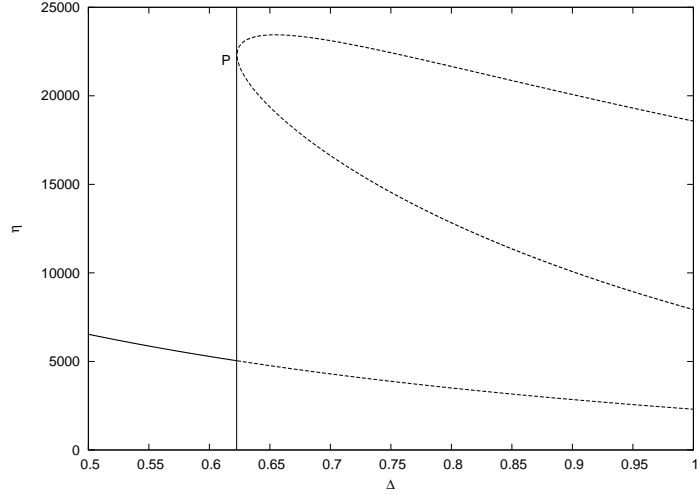}
\caption{\label{f1} \small{For the parameter values,
${\mathcal E} =0.0001$, $\gamma = 4/3$ and $\lambda = 1.7$, 
the roots of $\eta$ are divided into two classes, spanning across 
the range, $0.5 \leq \Delta \leq 1$. The segregation of the two 
classes is marked by the vertical line passing through the point, 
$\mathrm P$, which keeps shifting to the left of the plot, as 
$\lambda$ is increased. To the left of this line is the region where
$\mathcal{D} > 0$, with $\eta$ having one root only. On the right
is the region where $\mathcal{D} < 0$, with $\eta$ having three real
roots. The globally relevant root is given by the locus at the 
bottom, running continuously through both the regions. The cusp
on the right actually comprises the loci of the other two roots 
of $\eta$. These two distinct roots merge at $\mathrm P$, where 
$\Delta \simeq 0.62$.}}
\end{center}
\end{figure}

Quite evidently, the sign of $\mathcal D$ and, in consequence, 
the number of real
roots of $\eta$ will be determined by the value of the flow
parameters like $\mathcal E$, $\lambda$, $\gamma$ and $\Delta$
(on which $\sigma$ depends). For fixed representative values of three
of these parameters --- $\mathcal E$, $\lambda$ and $\gamma$ ---
the continuously varying dependence of $\eta$ on $\Delta$ has
been shown in Fig.~\ref{f1}. The two distinct regions in the plot,
corresponding to ${\mathcal D} >0$ and ${\mathcal D} <0$, show 
the existence of a single and triple real roots, respectively. 
The one root that runs continuously across both the regions is 
the one that will be relevant for all subsequent analysis. 

Once $\eta$ is known, it is a simple task thereafter to know 
the value of $\zeta$, $a$ and $b$. With this having been accomplished,
all the four roots of $r_{\mathrm c}$ from equation~(\ref{quartic})
can be identified by solving the two distinct quadratic equations,
\begin{equation}
\label{biquad}
r_{\mathrm c}^2 + A r_{\mathrm c} + \zeta = \pm \left(a r_{\mathrm c}
+ b \right) \,,
\end{equation}
corresponding to the two signs on the right hand side. Each of these 
roots will determine the spatial position of a critical point in the
flow space. Of these four roots, one is always real, and it is always
located within the event horizon of
the central black hole. Hence, it will be irrelevant for this 
study. Of the other three roots, one remains always real, while the
other two, depending on the flow parameters, can only be either both 
real or both complex (complex roots always occur in pairs). In the 
former case there will, therefore, be three real roots on physically 
relevant length scales outside the event horizon, and this situation  
will correspond to multitransonicity. 

\begin{figure}
\begin{center}
\includegraphics[scale=0.75, angle=0.0]{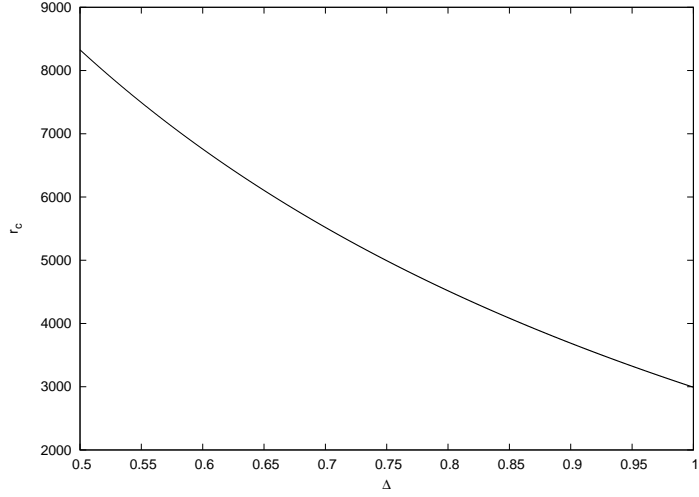}
\caption{\label{f2} \small{The position of the outermost critical 
point is shifted outwards when fractal properties emerge 
progressively stronger (as $\Delta$ decreases). The parameter 
values chosen are ${\mathcal E} = 0.0001$, $\gamma = 4/3$ and 
$\lambda = 1.7$.}}
\end{center}
\end{figure}

\begin{figure}
\begin{center}
\includegraphics[scale=0.75, angle=0.0]{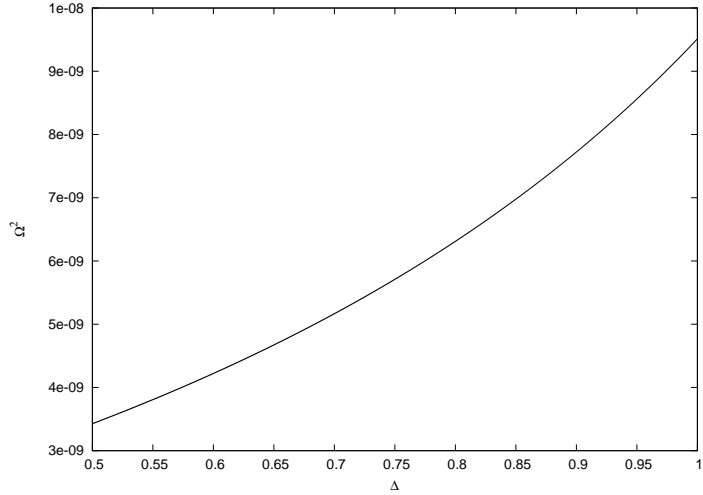}
\caption{\label{f3} \small{For the entire relevant range of 
$\Delta$, the outermost critical point behaves like a saddle 
point ($\Omega^2 > 0$). The parameter values remain the same at 
${\mathcal E} = 0.0001$, $\gamma = 4/3$ and $\lambda = 1.7$.
The saddle-type behaviour of this critical point makes 
transonicity possible in the flow. The fractal properties of 
the flow serve to enhance this effect.}} 
\end{center}
\end{figure}

Taking up the case of the root that always remains real at a given
physical
distance in the flow region, it can be seen from Fig.~\ref{f2} that 
as $\Delta$ decreases (i.e. as the fractal properties become more 
pronounced), the position of this critical point, $r_{\mathrm c}$, 
is monotonically shifted outwards. 
The nature of this critical point can be determined
from the plot in Fig.~\ref{f3}, which shows that $\Omega^2 >0$ for
the entire relevant range of $\Delta$. This can only mean that the
critical point implied by this particular root of $r_{\mathrm c}$, 
will always be a saddle point, and any solution passing through it, 
will necessarily be transonic. Practically speaking, all of this is
just how it should be. To make accretion a feasible process, there 
has to be at least one solution that will link the outer boundary
to the event horizon of the black hole. This can be achieved only
if the first critical point that a fluid element encounters while
travelling towards the accretor, after having started from an outer
boundary region, is a saddle point. So the steadfast existence of 
this particular saddle point (which is the outermost critical point
in the flow region), preserves the conditions necessary for transonicity
to develop. 
But why does the position of this saddle point get shifted outwards,
as the fractal properties become progressively more conspicuous? 
An earlier study on fractal flows in spherically symmetric geometry
has addressed this very question by arguing that a fractal medium 
behaves like an effectively more dilute continuous medium~\citep{rnr07},
and in being so, allows gravity to win over gas pressure resistance. 
The same
feature has appeared in this axisymmetric case. Here transonicity occurs 
when gravity wins over both the centrifugal effects and the pressure 
effects in an axisymmetric flow. The conditions
for transonicity are, therefore, more conducive when either of the
two effects is weak. A relatively more dilute gas will offer a feeble
pressure build-up 
against gravity, and so the gravitational pull can triumph
even at greater distances. It is for this reason that when the flow
is more fractal, that the transonic length scale is shifted outwards. 
The analogy with the spherically symmetric case is quite apt here.  

\begin{figure}
\begin{center}
\includegraphics[scale=0.75, angle=0.0]{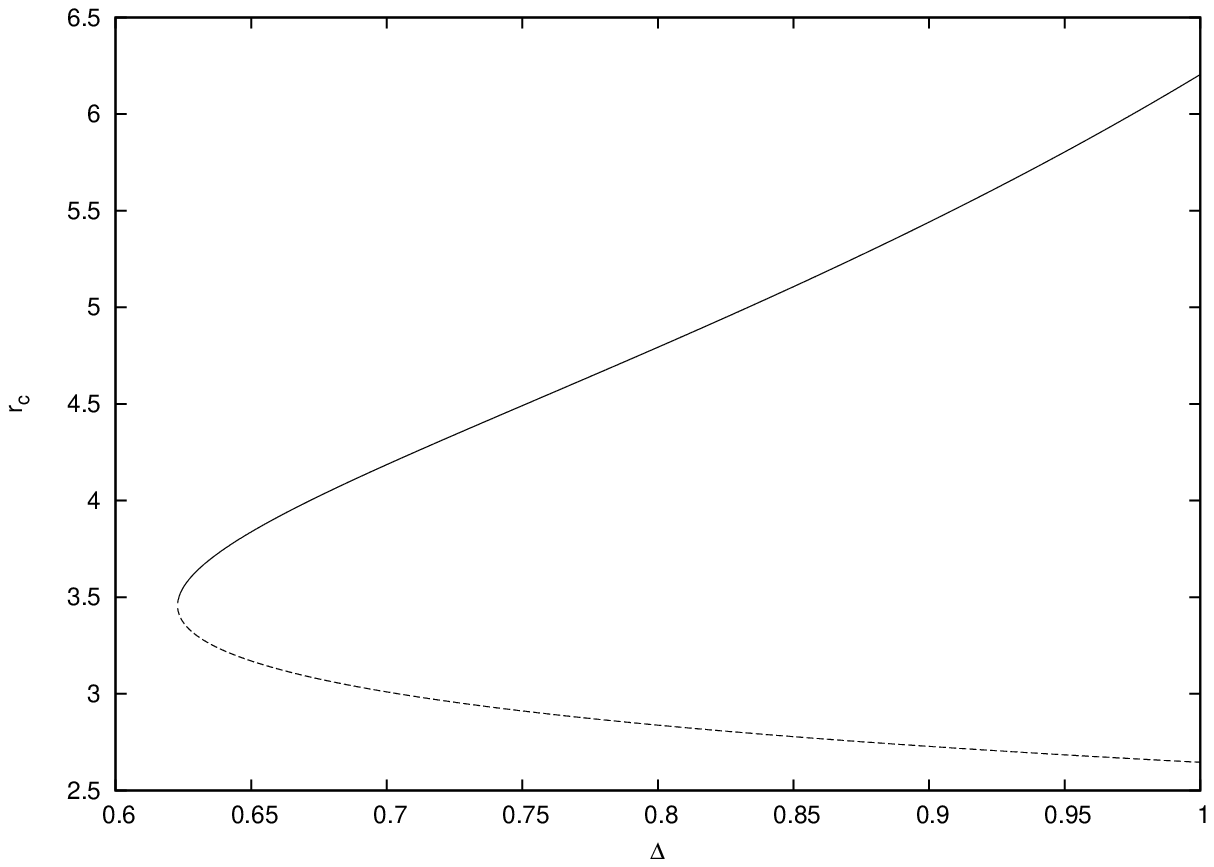}
\caption{\label{f4} \small{The upper arm of the cusp is the locus 
of the position of the middle critical point, for varying $\Delta$. 
The lower arm is likewise traced out by the innermost critical point. 
The two tracks coalesce at $r_{\mathrm c} \simeq 3.4$, when 
$\Delta \simeq 0.62$. For any lower value of $\Delta$, these two
critical points will cease to exist, with there being only one
saddle-type critical point, as shown in the earlier plots. The 
point of merging of the two critical points shifts to the left 
of the plot, if $\lambda$ is increased. As before, here the chosen 
parameter values are 
${\mathcal E} = 0.0001$, $\gamma = 4/3$ and $\lambda = 1.7$.}}
\end{center}
\end{figure}

\begin{figure}
\begin{center}
\includegraphics[scale=0.75, angle=0.0]{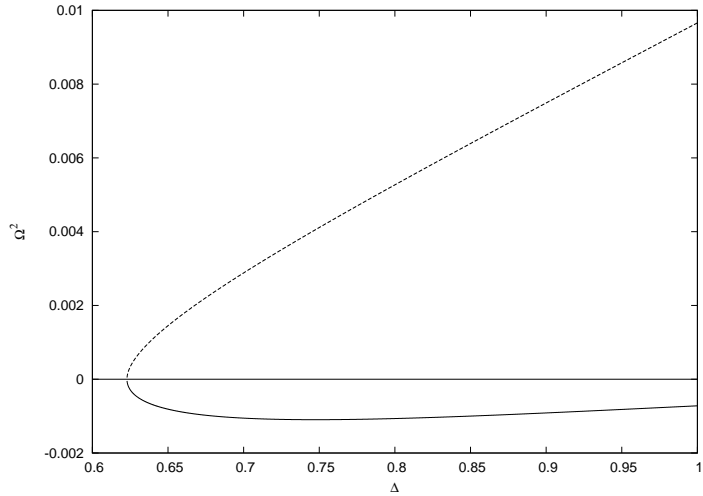}
\caption{\label{f5} \small{The nature of the innermost critical point 
is indicated by the upper arm of the cusp. This critical point 
always behaves like a saddle point ($\Omega^2 > 0$). The middle
critical point is always centre-type ($\Omega^2 < 0$). The two 
critical points merge when $\Omega^2 =0$ and $\Delta \simeq 0.62$. 
Therefore, multitransonicity is only possible for a limited range
if $\Delta$, in this case $0.62 \lesssim \Delta \leq 1$, when 
${\mathcal E} = 0.0001$, $\gamma = 4/3$ and $\lambda = 1.7$.}} 
\end{center}
\end{figure}

While this be the case for the outermost critical point, which is 
always a saddle point, to have an appreciation of how multitransonic
properties get affected by the fractal nature of the flow, it shall 
be necessary to turn to the other two critical points --- the 
innermost critical point and the middle critical point --- both 
situated very close to the accretor. These two critical points 
cannot exist without each other, and their collective behaviour 
bears this out, something that is shown in Fig.~\ref{f4}. The 
positions of the innermost critical point and the middle critical 
point approach each other as $\Delta$ is decreased, till they coalesce. 
Beyond this point, multitransonicity is not possible, and the 
accreting system will only have one saddle point left in its 
phase portrait (i.e. it will be monotransonic). The nature of 
both these critical points can be discerned from Fig.~\ref{f5}. 
The innermost critical point is always a saddle point, and the 
middle critical point is always a centre-type point. The two 
critical points merge under the condition, $\Omega^2 =0$, and this 
merging might be viewed as the mutual annihilation of a stable 
centre-type point (with $\Omega^2 <0$) and an unstable saddle 
point (with $\Omega^2 >0$) in the phase space. Significantly 
enough, the two critical points meet when the lower cut-off value 
of $\Delta \simeq 0.62$. This is the same value of $\Delta$, for 
which $\mathcal D$, as defined from equation~(\ref{discri}), vanishes, 
and the number of real roots of $\eta$ changes from one to three. 
In physical terms all of this will mean that if $\Delta$ is reduced 
further, i.e. if the flow becomes more fractal, it will rule out 
the  possibility of multitransonicity completely. 
Strongly fractal disc systems, therefore, would be devoid of
the multitransonic character usually associated with conserved
continuum disc accreting systems.  

So the general conclusion that one might draw regarding the
influence of the fractal features in the flow, is that the 
flow is made to behave like a more dilute continuum. While 
this helps in generating the transonic solution, in a somewhat 
paradoxical sense, the same fractal features turn out to be 
inimical to multitransonicity in the flow. These inferences   
have been made by studying a general polytropic flow, but 
arguably they will not be violated when the flow is to be 
considered as isothermal. In this limit also it is possible,
from equation~(\ref{critconiso}), to derive a quartic polynomial
just as in equation~(\ref{quartic}), with the coefficients $A$,
$B$, $C$ and $D$ being set down, respectively, as 
\begin{displaymath}
\label{coeffiso}
A = - \frac{2 c_{\mathrm s}^2 \left(2\sigma +1\right) +1}
{4 c_{\mathrm s}^2 \left(\sigma +1\right)} \,, \qquad \qquad
B = \frac{c_{\mathrm s}^2 \sigma + \lambda^2}{c_{\mathrm s}^2
\left(\sigma +1\right)} \,, \qquad \qquad 
C = - \frac{\lambda^2}
{c_{\mathrm s}^2\left(\sigma +1\right)} \,, \qquad \qquad 
D =  \frac{\lambda^2}{c_{\mathrm s}^2\left(\sigma +1\right)} \,.
\end{displaymath}
The rest of the mathematical analysis will remain the same, as
it has been for the polytropic flow. 

\section{Fractal scaling for the mass accretion rate}
\label{sec6}

With the knowledge that the critical conditions in the flow can 
be represented entirely in terms of flow parameters like 
$\dot{\mathcal{M}}$ (or $\dot{m}$) and $\mathcal{E}$, it now 
becomes possible to find a direct connection between the steady 
accretion rate and the 
mass of the accretor whose gravitational force field drives the 
fractal accretion flow. This should be particularly revealing for 
the case of the Newtonian potential, $\Phi (r)=-GMr^{-1}$. In this limit
it is can be shown from equation~(\ref{efixcrit}) that $r_{\mathrm c}$
is scaled as $GM{\mathcal E}^{-1}$ and $\lambda$ is scaled as 
$GM{\mathcal E}^{-1/2}$~\citep{rbcqg}. Using these scaling conditions 
in equation~(\ref{dotmfix}), one can find a scaling behaviour 
for the accretion flow rate, $\dot{m}$, going as 
\begin{equation}
\label{flowratescal}
\dot{m} \sim K^{-n} \left(GM\right)^{2\Delta}
{\mathcal E}^{n -\sigma} \,.
\end{equation}
This result is particularly interesting since it indirectly provides 
some insight about the nature of density clumpiness or the fractal 
structure of the accretion disc in a specific astrophysical situation. 

In the case of accreting pre-main-sequence (PMS) stars, it has been 
argued~\citep{padoan} that even if accretion takes place through the 
formation of 
accretion discs on relatively small length scales, on larger scales 
it can well be approximated as spherically symmetric~\citet{bon52}
or~\citet{bhoyle} accretion. This scheme is in good agreement with 
both numerical simulations and observational results~\citep{padoan}. 
Now pre-main-sequence stars are embedded in the interstellar 
medium (ISM), and numerous observational and numerical 
studies~\citep{crov,falg92,lang,elme,burk,fais,heith,kless,hill},
using a variety of techniques, have directly or indirectly suggested 
that the interstellar medium has a fractal structure over a 
wide range of scale lengths. So, it is more realistic to consider 
the case of accretion onto PMS stars as fractal accretion. In this
very context it has been shown by~\citet{roy07} that the numerical 
simulations and the currently available observational results 
pertaining to PMS star accretion rate, are consistent with the 
model of spherically symmetric fractal accretion. It has also 
been argued by~\citet{roy07} that the accretion rate is scaled as 
$\dot{m} \sim M^{D-1}$, where $D(<3)$ is the mass dimension of 
the surrounding ISM. So, if the proposed global model for accretion 
onto PMS stars --- spherically symmetric accretion at large distances 
from the star and disc accretion close to it --- is correct, then 
in the steady state, going by equation~(\ref{flowratescal}), one 
will have $2\Delta=D-1<2$. This will imply that the disc will also 
have a fractal structure, if the surrounding medium, from which the 
large scale accretion takes place, is fractal in nature.

\section{Dynamic aspects of the axisymmetric fractal flow}
\label{sec7}

The fractal nature of the accreting matter acts against 
multitransonicity. Nevertheless, this does not preclude 
transonicity itself. If anything, the static flow shows that
prominent fractal features will certainly bring about the existence 
of one saddle-type critical point in the phase portrait of the
flow. The overall appearance of a conspicuously sub-Keplerian
fractal flow will, therefore, be like that of a monotransonic
spherically symmetric~\citet{bon52} flow. This is fortunate in 
many respects, because the~\citet{bon52} solution has by now become
quite a well-understood and a regularly-invoked paradigm in accretion 
studies, and any
close convergence between it and the fractal disc flow, will 
pave the way for making insightful comparisons. 

The most significant of these is that there will at least be one 
continuous solution which will connect the outer boundary of 
the flow to the event horizon of the black hole accretor, 
in a process that will make black hole accretion realisable 
in its expected fashion. The steady state conditions will 
imply that this solution will pass through a saddle point,
which is known to be unstable~\citep{js99}. 
Any solution passing through this point will suffer from 
the problem of fine tuning the outer boundary condition with 
infinite precision~\citep{rb02}. These difficulties can be
avoided through the dynamics, and indeed in the case of fractal
spherically symmetric accretion, it has been shown~\citep{rnr07} 
that the stronger is the fractal nature of the flow, the more 
successful is the time-evolutionary drive towards the~\citet{bon52} 
solution. This
happens simply because a fractal medium translates equivalently
as a continuum with an effective lesser density, i.e. a fractal
flow can be construed as a more dilute continuum. Since a 
sizeable resistance against gravity-driven transonicity happens
due to the pressure build-up in the flow (which, through the 
polytropic relation, is connected to the local flow density), 
any dilution in the flowing medium will detract from the 
opposition against transonicity. Hence, this entire effect 
will enable an accreting solution to cross the sonic horizon 
smoothly, premised on the condition that this solution will 
also correspond to a minimum possible energy configuration, 
and, concomitantly, to the maximum possible inflow 
rate~\citep{bon52,gar79,rb02,rnr07}. 

This line of reasoning can be carried over fully to the case 
of inviscid axisymmetric accretion. As a matter of fact, 
in the context of inviscid thin-disc flows,~\citet{rbcqg} have 
already argued that transonicity is determined and governed
by the same dynamic and non-perturbative criteria, as they are 
for the~\citet{bon52} flow. The analogy with the~\citet{bon52} 
solution will be more so true for sub-Keplerian solutions with 
low angular momentum (having feeble centrifugal effects pitted
against gravity). Fractal features in this kind of disc configuration
will only serve to facilitate transonicity even further.  

As opposed to a completely non-perturbative approach to the 
question of time-dependence in the fractal disc flow (all of
which will require the mathematics of partial differential 
equations), it would also be worthwhile to study the properties
of the background stationary flow under the influence of a 
linearised time-dependent perturbative effect. This will yield
much information on the global stability of the flow solutions.
In order to achieve that, it will first be necessary to define 
a new physical variable, 
$\psi =\rho^{(\gamma +1)/2} v r^{\sigma}$, closely following a 
perturbative procedure prescribed by~\citet{pso80} and~\citet{td92}
for spherically symmetric flows, and applied successfully later to
thin disc flows~\citep{ray03,crd06,rbcqg}. It is quite obvious 
from the form of equations~(\ref{rhocon}) and~(\ref{con}), that the 
stationary value of $\psi$ will
be a global constant, $\psi_0$, which can be closely identified with the
matter flux rate, within a constant geometrical factor. A perturbation
prescription of the form $v(r,t) = v_0(r) + v^{\prime}(r,t)$ and
$\rho (r,t) = \rho_0 (r) + \rho^{\prime}(r,t)$, will give,
on linearising in the primed quantities,
\begin{equation}
\label{effprime}
\frac{\psi^{\prime}}{\psi_0} = \left(\frac{\gamma +1}{2}\right)
\frac{\rho^{\prime}}{\rho_0} + \frac{v^{\prime}}{v_0} ,
\end{equation}
with $\psi^{\prime}$ being a linearised time-dependent perturbation about
the constant matter inflow rate, $\psi_0$. It is significant that the
foregoing expression for $\psi^{\prime}$ is free of the explicit presence
of $\sigma$. Linearising in $\rho^{\prime}$ and $v^{\prime}$ about
$\rho_0$ and $v_0$, respectively, in both equations~(\ref{rhocon})
and~(\ref{fulleuler}), and expressing $\rho^{\prime}$ and $v^{\prime}$
separately in terms of $\psi^{\prime}$ only, will ultimately lead
to a linearised equation for the perturbation as
\begin{equation}
\label{tpert}
\frac{{\partial}^2 \psi^{\prime}}{\partial t^2}
+2\frac{\partial}{\partial r}
\left(v_0 \frac{\partial \psi^{\prime}}{\partial t}\right)+\frac{1}{v_0}
\frac{\partial}{\partial r}\left[ v_0 
\left(v_0^2- \beta^2 c_{{\mathrm{s}}0}^2\right)
\frac{\partial \psi^{\prime}}{\partial r}\right] = 0 ,
\end{equation}
which is an expression that is exactly the same as what can derived 
upon perturbing the stationary solutions of conserved axisymmetric 
inflows~\citep{ray03,crd06}. Another aspect of
equation~(\ref{tpert}) is that its form has no explicit dependence on
the potential that is driving the
flow. This is entirely to be expected, because the potential, being
independent of time, will appear only in the stationary background 
flow. Arguments regarding stability will, therefore, be more
dependent on the boundary conditions of the steady flow.
As the form of the equation of motion for the linearised perturbation
remains unchanged even for a flow in a fractal medium, and as the
physical boundary conditions are also not altered in this case, the
general conclusions reached earlier regarding non-fractal axisymmetric
flows~\citep{ray03}, can be extended here, and it
can be safely claimed that under all reasonable boundary conditions,
both the transonic and subsonic solutions will be stable. The only
difference that will arise will be due to a fractal scaling of the 
amplitude of the perturbation. Restricting
the perturbation to be like a high-frequency travelling wave, and 
carrying out a Wentzel--Kramers--Brillouin analysis on equation~(\ref{tpert}), will 
show that the time-averaged energy flux in the perturbation,
$\mathcal F$, goes as $\psi_0^{-1}$~\citep{ray03,crd06}. 
Since $\psi_0 \propto {\dot{m}}$, 
going by equation~(\ref{flowratescal}), one can argue that 
${\mathcal F} \sim M^{-2\Delta}$, something that gives the 
expected fractal scaling for the amplitude of the perturbation.  

\section{Concluding remarks}
\label{sec8}

Two salient facts have emerged out of the entire theoretical 
exercise carried out here: fractal flows can be equivalently
modelled as an effective continuum flow with a diminished local
density, and that fractal properties of the flow have an adverse
bearing on the question of multitransonicity. It is conceivable
that there should be some observational imprints of these two 
aspects of fractal flows. The mass accretion rate in an accretion
flow is intimately related to the local density field. It has 
already been argued here that the mass accretion rate undergoes
fractal scaling, and there are observational indications of such 
scaling for the particular case of accretion processes onto a 
PMS star.

In the context of multitransonicity, shocks are a very closely
related feature. Many earlier works have discussed the transition
of the flow from one transonic solution to the other via a 
shock~\citep{skc90,skc96}.
Now fractal flows have been shown to be detrimental to 
multitransonicity. So, if shocks do or do not arise in a fractal 
flow, then that should cast some light on the actual mechanism
behind shock formation in an accretion process. Again some 
observational signature of this should exist. 

The exact manner in which matter infall takes place
along the radius of an accretion disc, has been a subject of much
study~\citep{dlb69,pri81,ny94,fkr02}. The standard view is that 
viscosity (in some form) aids the outward transport of angular 
momentum~\citep{ss73,bh98}, and effects the formation of a Keplerian 
distribution of the accreting matter~\citep{pri81,fkr02}. What has
been shown here is that 
for a conserved low angular momentum sub-Keplerian disc,
a fractal structure in the accreting medium would drive the flow 
closer to the simple~\citet{bon52} accretion limit, and would 
therefore assist in the infall process onto a black hole. So
a combination of fractal behaviour and viscosity in the flow 
could possibly explain the observed infall rates. 

\section*{Acknowledgements}

This research has made use of NASA's Astrophysics Data System.
The authors express their gratitude to Jayanta K. Bhattacharjee,
Jayaram N. Chengalur, Tapas K. Das and an anonymous referee
for some helpful comments.

\bsp

\label{lastpage}

\begin{thebibliography}{99}

\bibitem[\protect\citeauthoryear{Abraham et al.}{2006}]{abd06}
Abraham H., Bili\'c N., Das T. K.,  2006, Classical and Quantum
Gravity, 23, 2371

\bibitem[\protect\citeauthoryear{Abramowicz \& Kato}{1989}]{ak89}
Abramowicz M. A., Kato S.,  1989, ApJ, 336, 304

\bibitem[\protect\citeauthoryear{Abramowicz \& Zurek}{1981}]{az81}
Abramowicz M. A., Zurek W. H.  1981, ApJ, 246, 314

\bibitem[\protect\citeauthoryear{Afshordi \& Paczy\'nski}{2003}]{ap03}
Afshordi N., Paczy\'nski B.,  2003, ApJ, 592, 354

\bibitem[\protect\citeauthoryear{Artemova et al.}{1996}]{abn96}
Artemova I. V., Bj\"ornsson G., Novikov I. D.,  1996, ApJ, 461, 565 

\bibitem[\protect\citeauthoryear{Balbus \& Hawley}{1998}]{bh98}
Balbus S. A., Hawley J. F.,  1998, Reviews of Modern Physics, 70, 1

\bibitem[\protect\citeauthoryear{Barai et al.}{2004}]{bdw04}
Barai P., Das T. K., Wiita P. J.,  2004, ApJ, 613, L49

\bibitem[\protect\citeauthoryear{Bohr et al.}{1993}]{bdp93}
Bohr T., Dimon P., Putkaradze V.,  1993, Journal of Fluid 
Mechanics, 254, 635

\bibitem[\protect\citeauthoryear{Bondi}{1952}]{bon52}
Bondi H.,  1952, MNRAS, 112, 195

\bibitem[\protect\citeauthoryear{Bondi \& Hoyle}{1944}]{bhoyle} 
Bondi H., Hoyle F.,  1944, MNRAS, 104, 273

\bibitem[\protect\citeauthoryear{Burkert et al.}{1997}]{burk} 
Burkert A., Bate M. R., Bodenheimer P.,  1997, MNRAS, 289, 497

\bibitem[\protect\citeauthoryear{Chakrabarti}{1989}]{c89}
Chakrabarti S. K.,  1989, ApJ, 347, 365

\bibitem[\protect\citeauthoryear{Chakrabarti}{1990}]{skc90}
Chakrabarti S. K.,  1990, Theory of Transonic Astrophysical
Flows, World Scientific, Singapore

\bibitem[\protect\citeauthoryear{Chakrabarti}{1996}]{skc96}
Chakrabarti S. K.,  1996, Physics Reports, 266, 229

\bibitem[\protect\citeauthoryear{Chandrasekhar}{1939}]{sc39}
Chandrasekhar S.,  1939, An Introduction to the Study of Stellar
Structure, The University of Chicago Press, Chicago

\bibitem[\protect\citeauthoryear{Chandrasekhar}{1981}]{sc81}
Chandrasekhar, S.,  1981, Hydrodynamic and Hydromagnetic Stability,
Dover Publications, New York

\bibitem[\protect\citeauthoryear{Chaudhury et al.}{2006}]{crd06}
Chaudhury S., Ray A. K., Das T. K.,  2006, MNRAS, 373, 146

\bibitem[\protect\citeauthoryear{Crovisier et al.}{1985}]{crov} 
Crovisier J., Dickey J. M., Kaz\`{e}s I.,  1985, A\&A, 146, 223

\bibitem[\protect\citeauthoryear{Das}{2002}]{das02}
Das T. K.,  2002, ApJ, 577, 880

\bibitem[\protect\citeauthoryear{Das}{2004}]{das04}
Das T. K.,  2004, MNRAS, 349, 375

\bibitem[\protect\citeauthoryear{Das et al.}{2007}]{dbd06}
Das T. K., Bili\'c N., Dasgupta S.,  2007, JCAP, 06, 009

\bibitem[\protect\citeauthoryear{Das et al.}{2003}]{dpm03}
Das T. K., Pendharkar J. K., Mitra S.,  2003, ApJ, 592, 1078

\bibitem[\protect\citeauthoryear{Elmegreen \& Falgarone}{1996}]{elme} 
Elmegreen B. G., Falgarone E.,  1996, ApJ, 471, 816

\bibitem[\protect\citeauthoryear{Faison et al.}{1998}]{fais} 
Faison M. D., Goss W. M., Diamond P. J., Taylor G. B.,  1998, 
AJ, 116, 2916

\bibitem[\protect\citeauthoryear{Falgarone et al.}{1992}]{falg92} 
Falgarone E., Puget J.-L., Perault M.,  1992, A\&A, 257, 715

\bibitem[\protect\citeauthoryear{Frank et al.}{2002}]{fkr02} 
Frank J., King A., Raine D.,  2002, Accretion Power in 
Astrophysics, Cambridge University Press, Cambridge

\bibitem[\protect\citeauthoryear{Fukue}{1987}]{fuk87}
Fukue J.,  1987, PASJ, 39, 309

\bibitem[\protect\citeauthoryear{Garlick}{1979}]{gar79}
Garlick A. R.,  1979, A\&A, 73, 171

\bibitem[\protect\citeauthoryear{Goswami et al.}{2007}]{gkrd07}
Goswami S., Khan S. N., Ray A. K., Das T. K.,  2007, MNRAS, 378, 1407

\bibitem[\protect\citeauthoryear{Heithausen et al.}{1998}]{heith} 
Heithausen A., Bensch F., Stutzki J., Falgarone E., 
Panis, J. F.,  1998, A\&A, 331, L65

\bibitem[\protect\citeauthoryear{Hill et al.}{2005}]{hill} 
Hill A. S., Stinebring D. R., Asplund C. T., Berkwick D. E., 
Everett W. B., Hinkel N. R.,  2005, ApJ, 619, L171

\bibitem[\protect\citeauthoryear{Jordan \& Smith}{1999}]{js99}
Jordan D. W., Smith P.,  1999, Nonlinear Ordinary Differential
Equations, Oxford University Press, Oxford

\bibitem[\protect\citeauthoryear{Kafatos \& Yang}{1994}]{ky94}
Kafatos M., Yang R. X.,  1994, MNRAS, 268, 925

\bibitem[\protect\citeauthoryear{Klessen et al.}{1998}]{kless} 
Klessen R. S., Burkert A., Bate M. R.,  1998, ApJ, 501, L205

\bibitem[\protect\citeauthoryear{Landau \& Lifshitz}{1987}]{ll87}
Landau L. D., Lifshitz E. M.,  1987, Fluid Mechanics,
Butterworth-Heinemann, Oxford

\bibitem[\protect\citeauthoryear{Langer et al.}{1995}]{lang} 
Langer W. D., Velusamy T., Kuiper T. B. H., Levin S., Olsen E., 
Migenes V.,  1995, ApJ, 453, 293

\bibitem[\protect\citeauthoryear{Larson}{1981}]{lars} 
Larson R. B.,  1981, MNRAS, 194, 809

\bibitem[\protect\citeauthoryear{Liang \& Thomson}{1980}]{lt80}
Liang E. P. T., Thomson K. A.,  1980, ApJ, 240, 271

\bibitem[\protect\citeauthoryear{Lu et al.}{1997}]{lyyy97}
Lu J. F., Yu K. N., Yuan F., Young E. C. M.,  1997, A\&A, 321, 665

\bibitem[\protect\citeauthoryear{Lynden-Bell}{1969}]{dlb69}
Lynden-Bell D.,  1969, Nature, 223, 690

\bibitem[\protect\citeauthoryear{Mandal et al.}{2007}]{mrd07}
Mandal I., Ray A. K., Das T. K.,  2007, 378, 1400

\bibitem[\protect\citeauthoryear{Mandelbrot}{1983}]{mand} 
Mandelbrot B.,  1983, The Fractal Geometry of Nature, W. H. Freeman, 
New York

\bibitem[\protect\citeauthoryear{Matsumoto et al.}{1984}]{mkfo84}
Matsumoto R., Kato S., Fukue J., Okazaki A. T.,  1984, PASJ, 36, 71

\bibitem[\protect\citeauthoryear{Molteni et al.}{1996}]{msc96}
Molteni D., Sponholz H., Chakrabarti S. K.,  1996, ApJ, 457, 805

\bibitem[\protect\citeauthoryear{Muchotrzeb-Czerny}{1986}]{mc86}
Muchotrzeb-Czerny B.,  1986, Acta Astronomica, 36, 1

\bibitem[\protect\citeauthoryear{Nakayama \& Fukue}{1989}]{nf89}
Nakayama K., Fukue J.,  1989, PASJ, 41, 271

\bibitem[\protect\citeauthoryear{Narayan \& Yi}{1994}]{ny94}
Narayan R., Yi I.,  1994, ApJ, 428, L13

\bibitem[\protect\citeauthoryear{Nowak \& Wagoner}{1991}]{nw91}
Nowak A. M., Wagoner R. V.,  1991, ApJ, 378, 656 

\bibitem[\protect\citeauthoryear{Paczy\'nski \& Wiita}{1980}]{pw80}
Paczy\'nski B., Wiita P. J.,  1980, A\&A, 88, 23

\bibitem[\protect\citeauthoryear{Padoan et al.}{2005}]{padoan} 
Padoan P., Kritsuk A., Norman M. L., Nordlund \AA.,  2005, ApJ, 622, L61

\bibitem[\protect\citeauthoryear{Pariev}{1996}]{par96}
Pariev V. I.,  1996, MNRAS, 283, 1264

\bibitem[\protect\citeauthoryear{Petterson et al.}{1980}]{pso80}
Petterson J. A., Silk J., Ostriker J. P.,  1980, MNRAS, 191, 571

\bibitem[\protect\citeauthoryear{Pringle}{1981}]{pri81}
Pringle, J. E.,  1981, ARA\&A, 19, 137

\bibitem[\protect\citeauthoryear{Ray}{2003}]{ray03}
Ray A. K.,  2003, MNRAS, 344, 83

\bibitem[\protect\citeauthoryear{Ray \& Bhattacharjee}{2002}]{rb02}
Ray A. K., Bhattacharjee J. K.,  2002, Phys. Rev. E, 66, 066303

\bibitem[\protect\citeauthoryear{Ray \& Bhattacharjee}{2007}]{rbcqg}
Ray A. K., Bhattacharjee J. K.,  2007, Classical and Quantum Gravity,
24, 1479

\bibitem[\protect\citeauthoryear{Ren et al.}{2003}]{ren} 
Ren F-Y., Liang J-R., Wang X-T., Qiu W-Y.,  2003, 
Chaos, Solitons and Fractals, 16, 107 

\bibitem[\protect\citeauthoryear{Roy}{2007}]{roy07}
Roy N.,  2007, MNRAS, 378, L34

\bibitem[\protect\citeauthoryear{Roy \& Ray}{2007}]{rnr07}
Roy N., Ray A. K.,  2007, MNRAS, 380, 733

\bibitem[\protect\citeauthoryear{Semelin \& Combes}{2000}]{seme} 
Semelin B., Combes F.,  2000, A\&A, 360, 1096

\bibitem[\protect\citeauthoryear{Shakura \& Sunyaev}{1973}]{ss73}
Shakura N. I., Sunyaev R. A.,  1973, A\&A, 24, 337

\bibitem[\protect\citeauthoryear{Strogatz}{1994}]{stro}
Strogatz S. H.,  1994, Nonlinear Dynamics and Chaos, Addison-Wesley
Publishing Company, Reading, MA

\bibitem[\protect\citeauthoryear{Tarasov}{2004}]{tara} 
Tarasov V. E.,  2004, Chaos, 14, 123

\bibitem[\protect\citeauthoryear{Theuns \& David}{1992}]{td92}
Theuns T., David M.,  1992, ApJ, 384, 587

\bibitem[\protect\citeauthoryear{Witten \& Sander}{1981}]{witsan}
Witten T. A., Sander L. M.,  1981, Phys. Rev. Lett., 47, 1400

\bibitem[\protect\citeauthoryear{Yang \& Kafatos}{1995}]{yk95}
Yang R. X., Kafatos M.,  1995, A\&A, 295, 238

\bibitem[\protect\citeauthoryear{Zaslavsky}{2002}]{zas} 
Zaslavsky G. M.,  2002, Physics Reports, 371, 461

\end{thebibliography}
\end{document}